\documentclass[fleqn,usenatbib,useAMS]{mnras}

\usepackage{graphicx}	% Including figure files
\usepackage{amsmath}	% Advanced maths commands
\usepackage{amssymb}	% Extra maths symbols
\usepackage{multicol}        % Multi-column entries in tables
\usepackage{bm}		% Bold maths symbols, including upright Greek
\usepackage{pdflscape}	% Landscape pages

\usepackage[T1]{fontenc}
\usepackage{ae,aecompl}

\usepackage{newtxtext,newtxmath}
\title[High z SFR from GRBs]{The Consequences of Gamma-ray Burst Jet Opening Angle Evolution on the Inferred Star Formation Rate}

\author[Lloyd-Ronning et al.]{Nicole M. Lloyd-Ronning,$^{1,2}$\thanks{E-mail: lloyd-ronning@lanl.gov}
Jarrett L. Johnson,$^{1}$
Aycin Aykutalp,$^{1}$
\newauthor
\\
$^{1}$Los Alamos National Lab, Los Alamos, NM, USA 87545\\
$^{2}$Department of Math, Enginnering, \& Science, University of New Mexico, 4000 University Dr., Los Alamos, NM, USA 87544\\
}

\pubyear{2020}

\begin{document}
\label{firstpage}
\pagerange{\pageref{firstpage}--\pageref{lastpage}}
\maketitle

\begin{abstract}
 Gamma-ray burst (GRB) data suggest that the jets from GRBs in the high redshift universe are more narrowly collimated than those at lower redshifts.  This implies that we detect relatively fewer long GRB progenitor systems (i.e. massive stars) at high redshifts, because a greater fraction of GRBs have their jets pointed away from us.  As a result, estimates of the star formation rate (from the GRB rate) at high redshifts may be diminished if this effect is not taken into account.  In this paper, we estimate the star formation rate (SFR) using the observed GRB rate, accounting for an evolving jet opening angle.  {\em We find that the SFR in the early universe ($z > 3$) can be up to an order of magnitude higher than the canonical estimates}, depending on the severity of beaming angle evolution and the fraction of stars that make long gamma-ray bursts. Additionally, we find an excess in the SFR at low redshifts, although this lessens when accounting for evolution of the beaming angle.   Finally, under the assumption that GRBs do in fact trace canonical forms of the cosmic SFR, we constrain the resulting fraction of stars that must produce GRBs, again accounting for jet beaming-angle evolution. We find this assumption suggests a high fraction of stars in the early universe producing GRBs - a result that may, in fact, support our initial assertion that GRBs {\em do not} trace canonical estimates of the SFR.   
  
\end{abstract}


%%%%%%%%%%%%%%%%%%%%%%%%%%%%%%%%%%%%%%%%%%%%%%%%%%

%%%%%%%%%%%%%%%%% BODY OF PAPER %%%%%%%%%%%%%%%%%%

% The MNRAS class isn't designed to include a table of contents, but for this document one is useful.
% I therefore have to do some kludging to make it work without masses of blank space.

%\section{To-Dos}
 
%\item{Get clarification of Beacom's first point.}

\section{Introduction}

 Understanding the global star formation rate (SFR) density is a key factor in understanding galaxy formation and evolution throughout the history of our Universe; additionally, it provides a cosmic census of the many diverse astronomical objects in our Universe (e.g., see \cite{HB06,Ken12,Krum14,MD14} and references therein). 
 %{\bf  See \cite{Shim09} first paragraph for a bunch of classic references on this.  Quote a few reviews like Kennicut 2012 and others?}  
 However, accurately determining the cosmological SFR is difficult for a number of reasons. Many of these issues have to do with the assumptions invoked when trying to connect observations to a physical star formation rate density, as well as accurately accounting for observational selection effects (see, e.g., \cite{HB06, MD14} for a discussion of these issues). Furthermore, observations themselves are limited - classic techniques using ultraviolet and far infrared measurements of galaxies are difficult at high redshifts; to get an accurate measurement of the star formation rate beyond a redshift of $3$ or so, multiple techniques must be employed.

  Because long gamma-ray bursts (lGRBs) are the most luminous explosions in the universe and because of definitive evidence of their association with massive star progenitors \citep{Gal98,Hjorth03,WB06,HB12}, they have long been suggested as tools with which to estimate the high redshift star formation rate \citep{LRFRR02,Jak05, Kist08,Yuk08,Kist09,WP10,RE12, TPT13,Lien14, PKK15, Char16, Le17, Kin19, EC20}.  However, there are a number of issues that make doing so difficult, essentially related to understanding exactly what types of stars and/or fractions of the global stellar population produce GRBs (including accounting for multiple GRB progenitors), and understanding how this relationship may change over cosmic time.  In addition, the distribution of the GRB beaming angle plays an important role in relating the GRB rate to the SFR. And - finally and importantly - observational selection effects in the detection of high redshift GRBs  must be taken into account.    \\ 
  
  Recently, \cite{LR19,LR20} examined a large sample of lGRBs with redshifts ($z$, in the range $0.1 \lesssim z \lesssim 5$) and found that the estimates of the jet opening angle, $\theta_{j}$, appear to be narrower at high redshifts than at low redshifts, with a best-fit functional form of $\theta_{j} \propto (1+z)^{-0.8 \pm 0.2}$ (we chose a power-law fit as straightforward way to quantify this relationship and its scatter).  \cite {LR20} argue that this may be a result of lower metallicity, higher mass (and therefore denser) stars at high redshifts collimating the GRB jet more, compared to less dense stars at lower redshifts. Several recent studies support this framework - e.g., \cite{Klen20} show that low metallicity leads to more compact stars, while \cite{Chrus2020} show a higher rate of metal-poor star formation at high redshift, leading to a top-heavy IMF. \cite{Shard2020} show that the presence of magnetic fields can suppress fragmentation in the early universe, leading to a top-heavy IMF at higher redshifts.  Additionally, low metallicity stars at high redshifts undergo less mass (and angular momentum) loss, and therefore may rotate more rapidly. This may have an effect on the jet collimation, potentially leading to more collimated jets at high redshift (for example, for a magnetically launched jet \citep{BZ77}, the angular momentum and magnetic field of the central engine may play a role in the degree of collimation of the jet (Hurtado, in prep)).  
  
  Regardless of the physical origin of the jet angle-redshift anti-correlation, a consequence of this relationship is that there exists a smaller fraction of {\em observable} GRB jets at high redshift, compared to those at lower redshifts.  In other words, because of the narrower collimation at high redshifts, there will be a higher fraction of GRBs with jets pointed away from Earth.  This leads to a higher density of GRB progenitors at high redshift than we would infer if we use a constant, non-evolving jet opening angle. This effect must be taken into account when using GRBs to estimate the high redshift star formation rate.
  
  \begin{figure}
    \centering
    \includegraphics[width=3.0in]{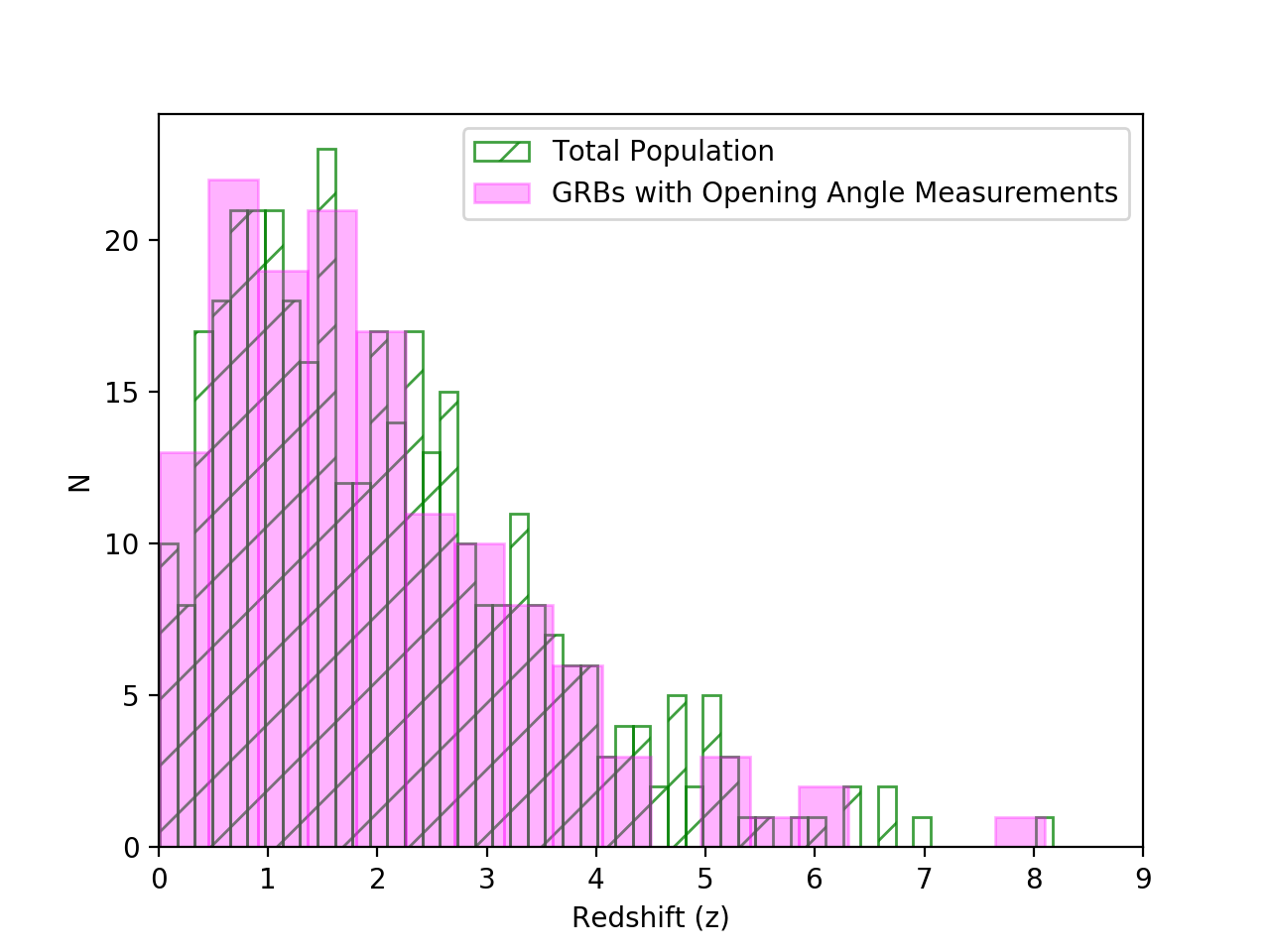}
    \caption{Redshift distribution of GRBs with jet opening angle distributions (magenta) compared to the entire population  There is no statistically significant difference between the shapes of the two distributions.}
    \label{fig:Reddist}
\end{figure}
  To estimate the star formation rate from the GRB rate, one must assume something about the fraction of stars that produce GRBs, and whether this fraction evolves through cosmic time (e.g. see \cite{Kist08,Yuk08,Kist09} for a straightforward summary of this issue).   Alternatively, one can assume a one-to-one correspondence between the GRB rate and the SFR measured by other techniques, and then infer the fraction of stars that produce GRBs.  Once again, GRB jet beaming angle evolution will affect this result and must be accounted for.
  
  In this paper, we examine both approaches with the novel addition of accounting for beaming angle evolution through cosmic time.  Our aim is twofold: 1) assuming the fraction of stars that produce lGRBs, {\em estimate the star formation rate} from the GRB rate accounting for the fact that lGRB beaming angle appears to evolve with redshift, and 2) under the assumption that lGRBs trace previously determined parameterizations of the global star formation rate, {\em estimate the fraction of stars that must produce lGRBs} in order to be consistent with the GRB rate (again, accounting for jet beaming angle evolution).   
  
  Our paper is organized as follows.  In \S 2, we summarize the data sample and results of \cite{LR19, LR20}, who showed lGRBs appear to exhibit cosmic beaming angle evolution, with higher redshift lGRBs more narrowly beamed than low redshift ones. In \S 3, we describe the method used to estimate the star formation rate and/or fraction of stars that are progenitors for lGRBs, based on the methods described in \cite{Kist08, Yuk08, Kist09}, and present our results.  We show that lGRB beaming angle evolution leads to a star formation rate at high redshifts that is higher than canonical estimates \citep{MD14} for both a constant and evolving fraction of stars that produce GRBs.   Alternatively, under the assumption that the lGRB rate density follows the \cite{MD14} star formation rate density, we calculate the {\em inferred} fraction of stars that make GRBs (and its evolution). These results indicate that a higher fraction of stars produce GRBs at both low ($(1+z) <3$) and high (($(1+z)>3$) redshifts relative to the peak of star formation - a counter-intuitive result that we argue may emphasize the inaccuracy of assuming that GRBs trace the global SFR.  Our conclusions are summarized in \S  4.
  
  \begin{figure*}
    \centering
    \includegraphics[width=3.5in]{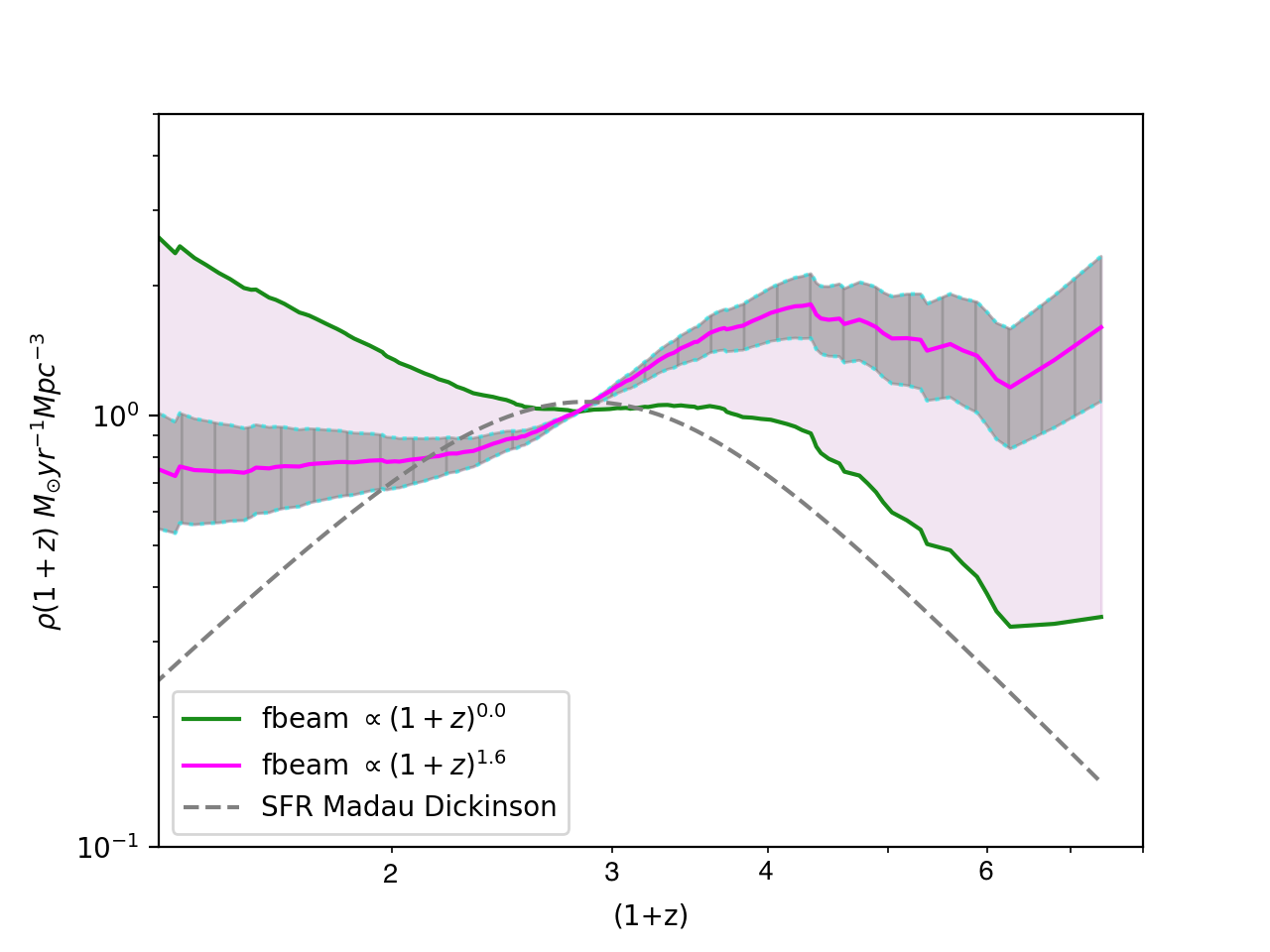}\includegraphics[width=3.5in]{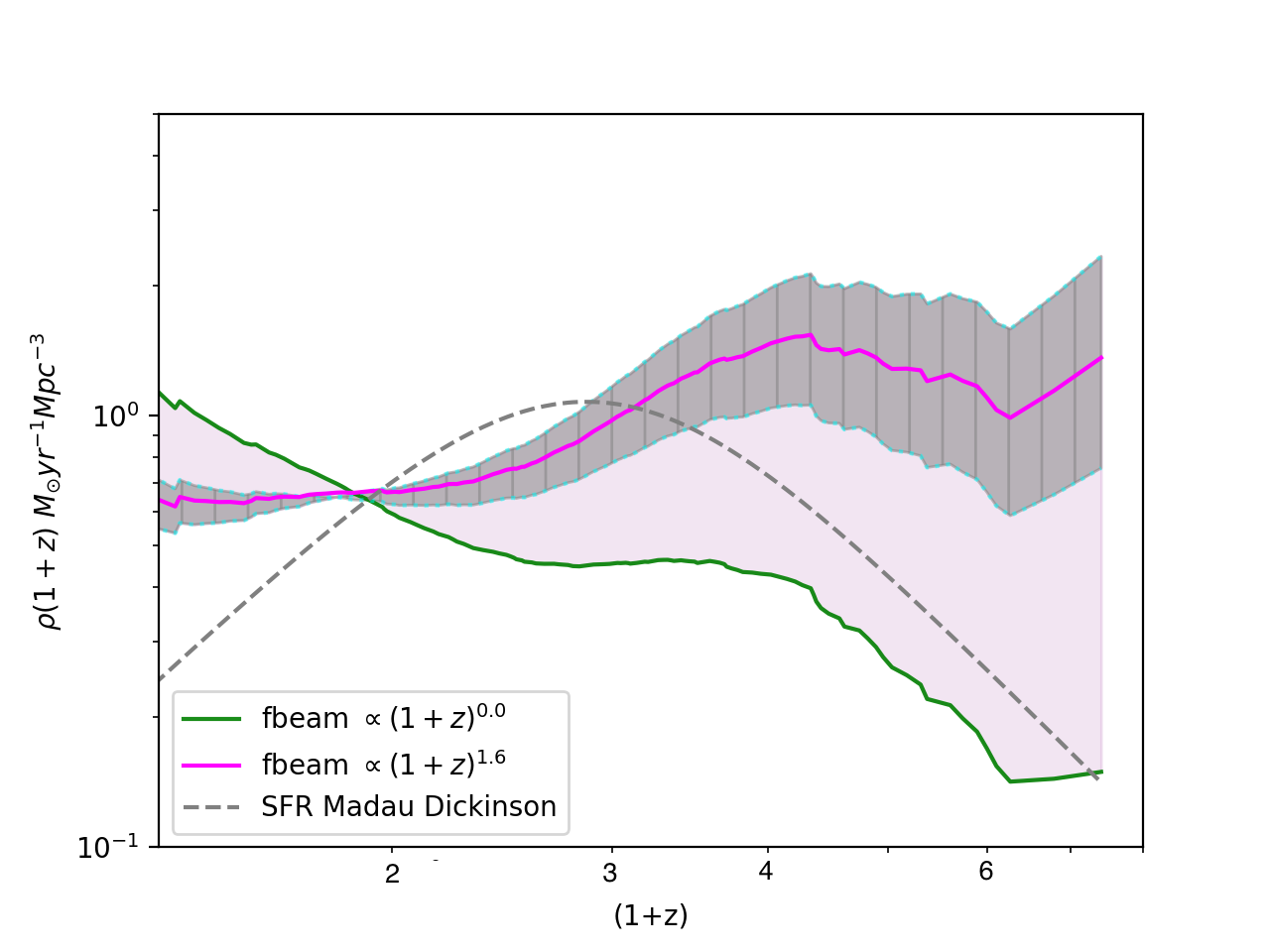}
    \caption{{\bf Left Panel:}Star formation rate density $\rho(1+z)$ as a function of redshift $(1+z)$ assuming a constant fraction of stars produce GRBs, and accounting for beaming angle evolution, according to the best fit to the data, $\theta_{j} \propto (1+z)^{-0.8 \pm 0.2}$ (with the gray region denoting the error on the fit). The green line shows the inferred SFR assuming no beaming angle evolution.  Curves are normalized to the MD14 star formation rate at a peak at $(1+z) = 3$. {\bf Right Panel:} Same as left panel but curves are normalized to the MD14 star formation rate at a redshift $(1+z) = 2$. }
    \label{fig:SFR}
\end{figure*}
 
 \section{Data}
   Our data sample is described in detail in \cite{LR19} and \cite{LR20}, who use data compiled in \cite{Wang2019}; this latter reference contains all publicly available observations of $6289$ gamma-ray bursts from 1991 to 2016. For the 376 GRBs with redshifts (and therefore isotropic energy) estimates, \cite{LR19} found that certain intrinsic long gamma-ray burst (lGRB) properties appear to evolve with redshift, even when accounting for Malmquist-type biases and selection effects in the observed data.  However, in the hundred or so bursts where jet opening angle estimates are available and for which one can compute beaming angle corrected (that is, the actual emitted) gamma-ray energy and luminosity, \cite{LR19} found these variables (i.e. gamma-ray luminosity and emitted energy) are {\em not} correlated with redshift.  This suggests  that jet opening angle {\em is}, and indeed they found a significant anti-correlation between jet opening angle and redshift, with a functional form $\theta_{j} \propto (1+z)^{-0.8 \pm 0.2}$.  Such an anti-correlation between jet angle and redshift was originally suggested in \cite{LRFRR02} (e.g. see their section 5.1.2; they suggested the faster rotation of stars at high redshift could be consistent with lower mass (and angular momentum) loss due to lower metallicity). Observational evidence for this anti-correlation has also been put forth by \cite{Lu12} and \cite{Las14,Las18,Las18b}.  An explanation for this correlation in terms of collimation by a massive star stellar envelope is given in \cite{LR20}. 
   
   \subsection{The Role of Selection Effects}
   As mentioned above, the analysis of \cite{LR19} accounts for gamma-ray flux-limit selection effects in the data.  However, we might ask what other types of selection effects could potentially contribute to the $\theta_{j}-(1+z)$ anti-correlation.  In this case, it is important to consider jet opening angle estimate techniques and whether there is a selection against higher opening angles at larger redshifts.  Indeed since opening angles are measured by breaks in afterglow light curves, when the relativistic beaming angle $1/\Gamma$ reaches the physical ``edge'' of the jet, larger opening angles cannot be detected until later times (for a given $\Gamma$ for the outflow) and the afterglow may have faded below detector sensitivity by that point.   
   
   We have explored these issues in \cite{LR19,LR20}  (see also the recent paper by \cite{Le20} who look at biases in redshift distributions and jet opening angles between different subsets of GRB data.).  In particular, in \cite{LR20} we implemented a strong artificial truncation in the $\theta_{j}-(1+z)$ plane, which mimics a selection against large opening angles at high redshifts.  Even when accounting for this selection bias using established non-parametric statistical techniques \citep{EP92,ep98}, we find there is still a significant anti-correlation between $\theta_{j}$ and $(1+z)$.  We also note that the redshift distribution of GRBs with jet opening angle estimates is not different from that of the entire sample of GRBs. A Kolmogorov-Smirnov (KS) test comparing the two distributions gives a p value of 0.64 that they are drawn from the same parent distribution - in other words, the redshift distributions of GRBs with jet opening angle measurements and the entire sample of GRBs are statistically the same (histograms of the two samples are shown in Figure~\ref{fig:Reddist}). Because there is roughly same relative fraction of jet opening angle measurements at low and high redshifts ($\sim 1/3$), this suggests we may not be missing a large fraction of large opening angle GRBs at high redshift and that the anti-correlation between jet opening angle and redshift may indeed have a physical origin.  
   
   %{\em Use some of the stuff we discussed in our previous paper, where we have a whole section on this. Jarrett's comment: ``If there are not a decreasing number of jet opening angles at higher redshifts, we may not need to worry as much about this effect - i.e. if no decreasing fraction of measurements at high z, no reason to suspect missing a bunch of higher $\theta_{j}$ lGRBs unless we expect the rate of lGRBs at high z to be much higher than low z.''  {\bf See the histogram figures at end of this draft.  A KS test on the redshift distributions of all long GRBs vs. those with jet opening angles gives that they are from the same parent distribution (p value of 0.64).  That means roughly same relative fraction of jet opening angle measurements at low and high redshift - this fraction is roughly $1/3$. Shouldn't be missing a set at high z, therefore.}}

   In what follows, we assume the jet opening angle-redshift anti-correlation is physical, and explore how this relationship can affect estimates of the high redshift star formation rate.

 \section{Results}
   Because of the strong evidence that lGRBs are associated with the deaths of massive stars (e.g. see \cite{WB06, HB12} for summaries), and because they are so luminous and can be detected to such high redshifts, many authors have attempted to use GRBs to estimate the high redshift star formation rate \citep{LRFRR02,Kist08,Yuk08,Kist09,WP10,RE12, TPT13, Lien14, PKK15, Kin19}.  However, as mentioned in the introduction, there are a number of complicating issues in understanding exactly how lGRBs track or trace the global star formation rate. One must address a host of issues in accurately determining the lGRB rate - flux sensitivity selection effects and other observational biases must be accounted for (e.g. \cite{LRFRR02, PKK15, LR19}) to get an accurate measure of the true, underlying lGRB rate.   Additionally, because we will in general only observe a fraction $d\Omega/4\pi$ of those GRBs whose jets are directed toward us (where $d\Omega = 2 \pi (1-cos(\theta_{j}))$ is the jet solid angle), an understanding of the behavior of GRB beaming angle distribution is necessary to correct for the true underlying number of GRB progenitor systems.  Finally, we need to get a handle on the GRB progenitor system - exactly what fraction of stars make long gamma-ray bursts, and how does this fraction evolve as a function of redshift?
   
  \subsection{Obtaining the Star Formation Rate from the lGRB Rate} 
   There are several possible approaches to tackling this problem.  One straightforward approach is laid out in \cite{Kist08,Yuk08,Kist09}.  One can essentially parameterize the various unknowns mentioned above to estimate the star formation rate from the lGRB rate:
 { \large 
  \begin{equation}
     \dot{\rho}_{\rm SFR}(z) = (\dot{dN}/dz)({\rm f_{beam}(z)})  \left(\frac{(1+z)}{dV/dz} \right) \frac{1}{\epsilon(z)}
  \end{equation}  } \\
 
 \noindent where $\dot{dN}/dz$ is the true, underlying lGRB rate (accounting for the GRB luminosity function and detector trigger selection effects), $\epsilon(z)$ parameterizes the fraction of stars that make GRBs (and in principle can evolve with redshift), and $f_{beam}(z)$ is a factor ($>1$) that accounts for the number of GRBs missed due to beaming.  The factor $dV/dz$ is the cosmological volume element given by:
 \begin{equation}
\begin{split}
    dV/dz = & 4 \pi (\frac{c}{H_{o}})^{3} \bigg[\int_{1}^{1+z} \frac{d(1+z)}{\sqrt{\Omega_{\Lambda} + \Omega_{m}(1+z)^{3}}}\bigg]^{2} \\
    & \times \frac{1}{\sqrt{\Omega_{\Lambda} + \Omega_{m}(1+z)^{3}}}
\end{split}
\end{equation}

 \noindent where we use $\Omega_{m} = 0.286$, an $\Omega_{\Lambda} = 0.714$ and an $H_{o} = 69.6 \ {\rm km \ s^{-1} Mpc^{-1}}$.   \cite{LR19} describe how they obtained the differential rate distribution of long GRBs as a function of redshift, $\dot{dN}/dz$, using the non-parameteric methods of \cite{LB71,EP92,ep98}. In particular, this quantity reflects the underlying GRB rate distribution, accounting for observational selection effects.  We refer the reader to \cite{LR19} for a discussion of how this distribution is obtained.  The factor we focus on here is $f_{beam}(z)$.  In previous studies, this was assumed to be a constant.  The results of \cite{LR19,LR20}, however, suggest that this function evolves with redshift.  This factor - a number greater than one, which parameterizes the number of GRBs missed due to jets being pointed away from us - is proportional to the inverse of the solid angle of the jet.  Therefore, because the solid angle is proportional to $\theta_{j}^{2}$ for small jet opening angles, if the jet opening angle $\theta_{j}$ evolves as $(1+z)^{-\alpha}$, the function $f_{beam} \propto (1/\theta_{j}^{2}) \propto (1+z)^{2\alpha}$.
  
  \begin{figure*}
    \centering
    \includegraphics[width=3.5in]{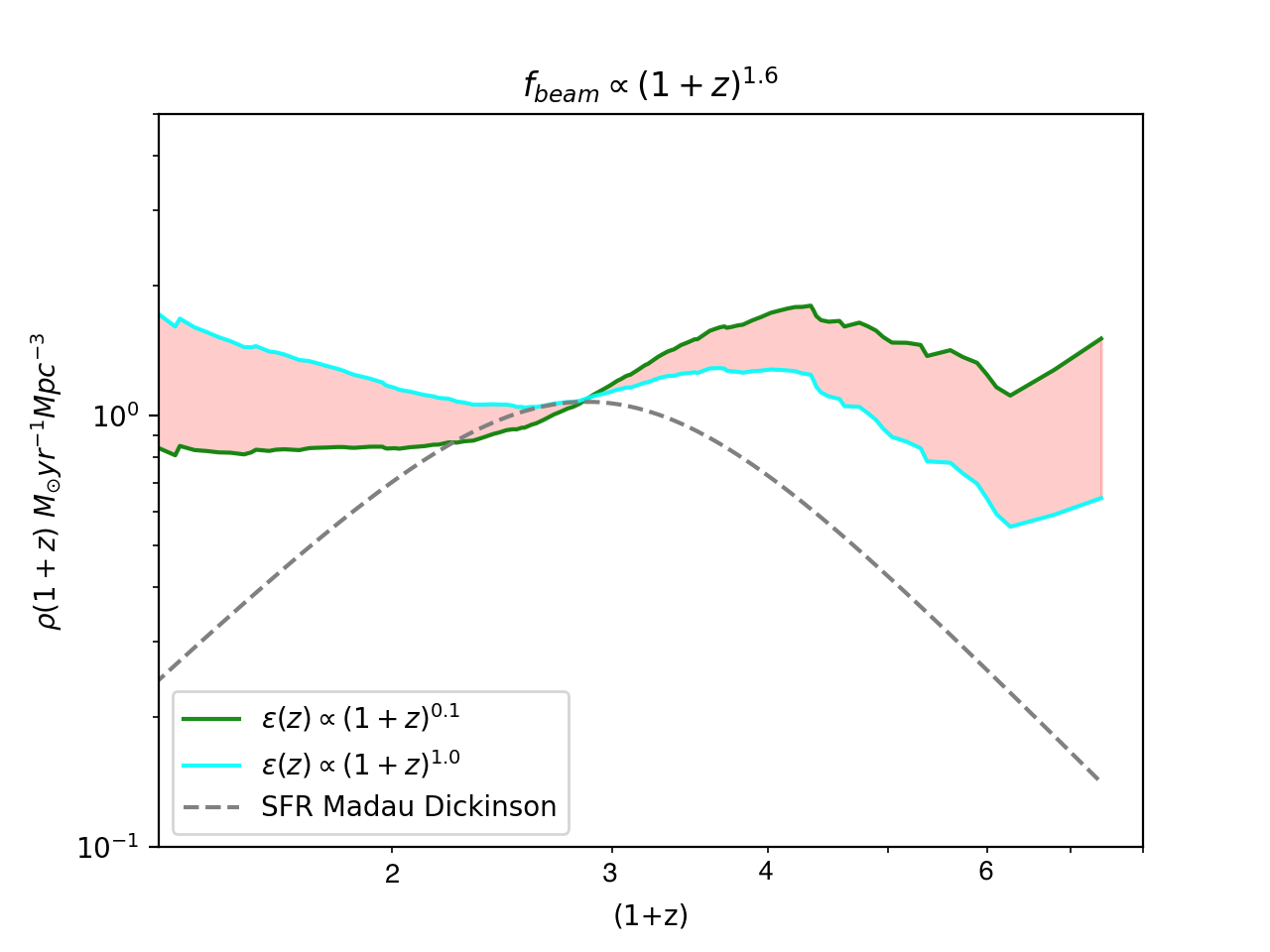}\includegraphics[width=3.5in]{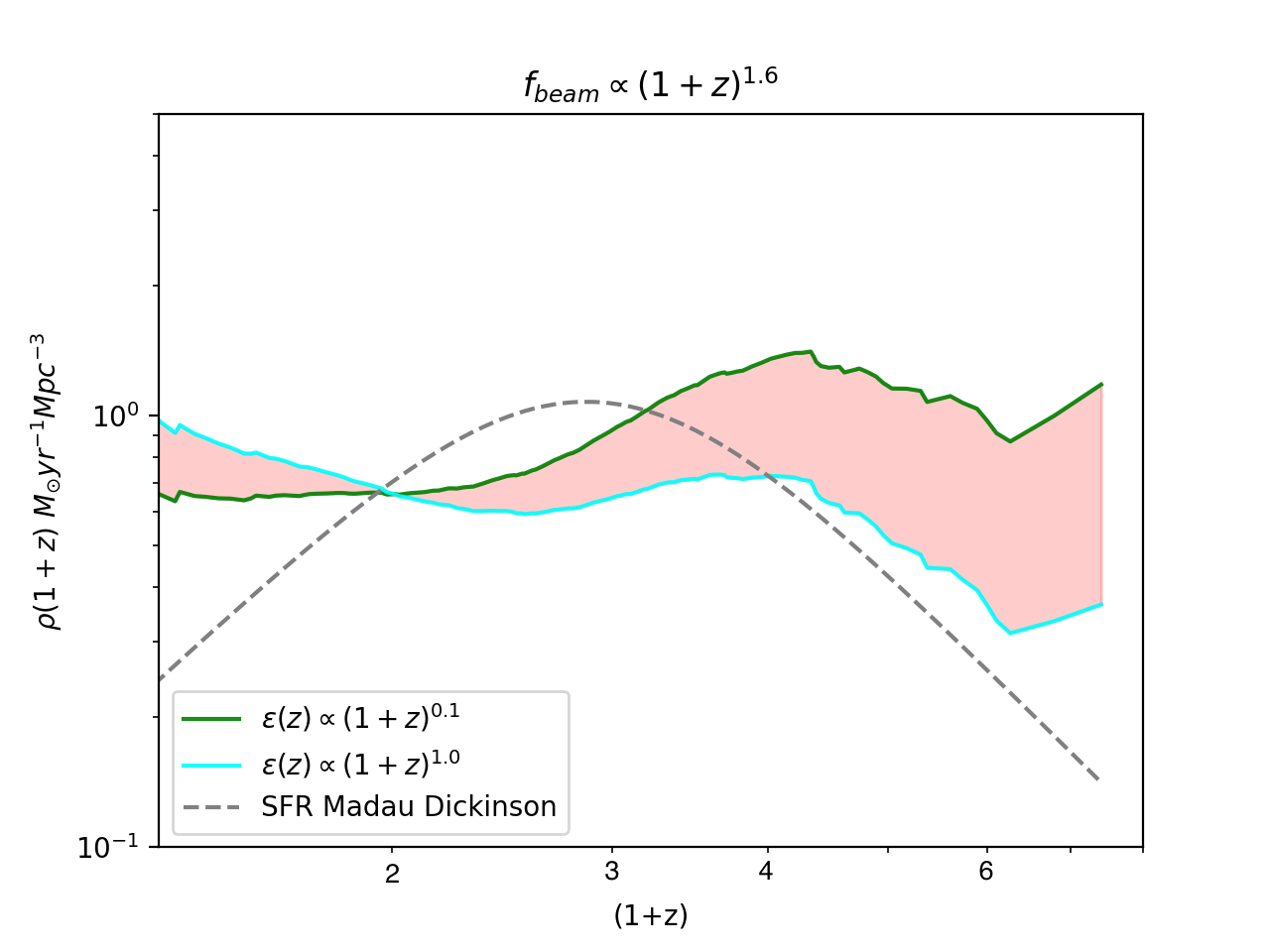}
    \caption{{\bf Left Panel:}Star formation rate density $\rho(1+z)$ as a function of redshift $(1+z)$ assuming the fraction of stars $\epsilon(1+z)$ that produce GRBs evolves with redshift, with $\epsilon(1+z) \propto (1+z)^{0.1}$ (green line) and $\epsilon(1+z) \propto (1+z)^{1.}$. (cyan line).  We take a beaming angle evolution of $f_{beam} \propto (1+z)^{1.6}$, consistent with the anti-correlation we find in the data between jet opening angle and redshift. Curves are normalized at the peak of the MD14 SFR.  {\bf Right Panel:} Same as left panel but with the curves normalized to the MD14 star formation rate at a redshift of $(1+z) \sim 2$.}
    \label{fig:SFRevolv}
\end{figure*}

  \cite{LR19} found $\alpha \sim 0.8$, which leads to $f_{beam}(z) \propto (1+z)^{1.6}$.   Figure~\ref{fig:SFR} shows the star formation rate derived from equation 1 above, given the functional form of beaming angle evolution seen in the data (magenta line, with the error indicated by the gray region). Here we have assumed that the fraction of stars $\epsilon(z)$, that produce lGRBs remains relatively constant throughout cosmic time. The green line in this figure shows the inferred SFR assuming no beaming angle evolution (but still a constant $\epsilon(z)$).  As expected, if lGRBs are more narrowly beamed in the high redshift universe, then - for a given fraction of stars that make lGRBs - there is a relatively higher star formation rate in the early universe.
  
  Because of the uncertainties in associating the lGRB rate with the global star formation rate, there is some freedom in how to normalize our star formation rate curves in Figure~\ref{fig:SFR}. One possibility is to normalize the star formation rate derived from the lGRB rate with that of the \cite{MD14} (hereafter, MD14) rate at a redshift of $(1+z) \approx 3$, where star formation appears to peak. This is shown in the left panel of Figure~\ref{fig:SFR}.  However, the star formation rate is better determined observationally at lower redshifts (see, e.g., Figure 1 of \cite{HB06}), and therefore normalizing our curves to the MD14 rate at $(1+z) \approx 2$ (or even lower) is also justifiable.  We show this normalization in the right panel of Figure~\ref{fig:SFR}. Note there appears to be an excess at low redshifts (particularly when beaming angle evolution is {\em not} taken into account), which we discuss further in \S 3.3 below.
  
  Regardless of normalization, Figure~\ref{fig:SFR} indicates that the shape or functional form of the SFR throughout cosmic time, as inferred from the GRB rate, is different from the MD14 rate (given a constant fraction of stars that make lGRBs).  In particular, when beaming evolution is accounted for, the peak of the SFR appears at redshifts of $z \sim 3$ (or higher) and there is a higher rate of SFR in the early universe than predicted by other estimates (on which the MD14 rate is based).

  Of course there is no reason to expect that the fraction of stars that make lGRBs should be constant throughout cosmic time.  Given the conditions of low metallicity (and, relatedly, high angular momentum) necessary to launch a GRB jet \citep{MW99,YL05,HMM05,Yoon06,WH06}, we might expect that a higher fraction of stars in the early universe make lGRBs compared to those in the lower redshift universe.  How exactly to parameterize or account for this is unclear, however.  In Figure~\ref{fig:SFRevolv} we show the star formation rate assuming two different functions for the evolution of the fraction of stars that make GRBs: $\epsilon(z) \propto (1+z)^{0.1}$ (green curve) and $\epsilon(z) \propto (1+z)^{1.0}$ (cyan curve). In these figures, we use a beaming evolution consistent with the relationship found in \cite{LR19,LR20}, $f_{beam} \propto (1+z)^{1.6}$. Our results indicate, again, that - whether or not the fraction of stars that make GRBs evolves through cosmic time - the SFR derived from the GRB rate is different from the MD14 rate, and higher at large redshifts, when beaming angle evolution is accounted for.

\subsection{The High Redshift Star Formation Rate} 
 Accounting for potential lGRB beaming angle evolution has a significant effect on the inferred high redshift star formation rate, leading to estimates that are up to an order of magnitude higher than the MD14 rate.   Interestingly, the peak of the inferred SFR (even without accounting for beaming angle evolution) appears to be around $(1+z) \sim 4$, compared with $(1+z) \sim 3$ of the MD14 rate. This may be a reflection of the lGRB rate tracing the evolution of a specific progenitor (e.g. low metallicity, massive stars) rather than the global stellar population.  In addition, our SFR curve is fairly flat from redshifts between $3.5 < (1+z) < 6$.  A similarly flat curve was found in the analyses of \cite{Kist08, PKK15, LR19}, without accounting for beaming angle evolution (although their inferred SFRs are flat between slightly different redshift ranges).\\
 
It is possible that the lGRB rate at high redshifts more closely follows galactic nuclear star formation, leading to a different redshift peak compared to MD14 rate. For example, \cite{HB06} suggest that the accretion of gas onto central supermassive black holes, triggered by mergers and/or interactions of galaxies, leads to starbursts (and active galactic nuclei (AGN) activity).  This AGN activity is expected to peak around $(1+z) \sim 4$ \citep{Miy15}, closer to the peak of the SFR we derive from the lGRB rate. Indeed numerical simulations have shown that the gravitational tidal torques excited during major mergers lead to rapid inflows of gas into the centers of galaxies \citep{BH96} which can be a mechanism to trigger starbursts in galaxies.  In addition, \cite{HQ10} find that AGN activity is more tightly coupled to nuclear star formation than the global star formation rate of a galaxy. This is also seen in numerical simulations of \cite{Ayk14, Ayk19}. Finally, \cite{HS10} found that in the X-ray irradiated case, fewer stars are formed but {\em with a higher initial masses}. Therefore, again, the lGRB rate may align more with this channel of star formation and will lead to an SFR peak that occurs earlier than the MD14 rate.

\subsection{On the Excess Rate at Low Redshifts}
The star formation rate we derive from the lGRB rate shows an excess at low redshifts compared to the MD14 rate.  The effect is less pronounced when we account for beaming angle evolution (but still there to some extent).  
We note that at very low redshifts (as $(1+z) \rightarrow 1$), the volume element (e.g. equation 2) goes to zero faster than the observed lGRB rate ($\dot{dN}/dz$) does,  and this causes the star formation rate in equation 1 to diverge at low redshifts. This effect comes into play around a redshift of $z \sim 0.3$; as a result, we show our results down to that limit, before the divergence becomes too severe (see also the discussion in \cite{LR19} of this issue). 

 However, even before this numerical effect comes into play, an excess at low redshifts appears to exist. This was also noted in \cite{PKK15, Yu15} and \cite{LR19}. We emphasize that these analyses account for the greater probability of detecting low luminosity GRBs at low redshifts (i.e. Malquist biases) through non-parametric statistical techniques that account for the GRB luminosity function (although we caution a single - albeit conservative - detector flux limit was used in our analysis; in reality the detector trigger criteria are more complicated). Another approach is to impose a minimum luminosity cutoff as in \cite{Kist08} - this will eliminate the excess of low luminosity GRBs at small redshifts (and as a result mitigate the excess in the inferred SFR at these redshifts).

 Again, this effect is more pronounced when beaming angle evolution is {\em not} accounted for.  Therefore, it may be that beaming angle evolution is stronger than we have estimated and the low redshift star formation rate in fact roughly matches the MD14 rate at low redshifts (as in the lower part of the gray region in Figure~\ref{fig:SFR}) (in this case, the high redshift star formation rate is then vastly larger than that of the MD14 rate).
 
 %Because the overall normalization of the SFR (derived from the observed GRB rate) is a free parameter, it could be that the curves should be normalized to the MD14 rate at even lower redshifts (say, $z \sim 0.1$, where the SFR is perhaps best determined).  However, we point out that although this would mitigate any ``excess'' at low redshifts, the SFR derived from the GRB rate still does not agree with that of the MD14 rate over a large range of redshifts.

 Another possibility for the mismatch at low redshifts could result from the array of progenitors that potentially contribute to lGRB rate \citep{Lev16}, which may be more pronounced at low redshifts. That is, there may exist a greater number lGRB progenitor systems that are  viable at lower redshifts.  For example, certain binary merger systems proposed for lGRBs - which require more cosmic time to form and merge - may play a larger role in the lGRB rate at low redshifts. Additionally, they do not necessarily need the low metallicity conditions required of single star progenitors \citep{MT20, Hao20}.  Meanwhile, single star progenitors may become less viable at low redshifts due to the higher metallicity and and accompanying higher mass loss  \citep{Chr20,PW20,Klen20, MT20}.

Finally, it may also be that the functional form of the parameterizations in equation 1 (particularly $f_{beam}(z)$ and $\epsilon(z)$) are not simple power laws, but are more complicated than what we have assumed.  We argue (here and in \cite{LR19,LR20}) that the data are reasonably parameterized by a power-law for $f_{beam}(z)$.  However, $\epsilon(z)$ could potentially be a very complicated function and indeed as we show below, when the GRB rate is assumed to follow the MD14 star formation rate, an interesting function for $\epsilon(z)$ emerges.

\begin{figure}
    \centering
    \includegraphics[width=3.5in]{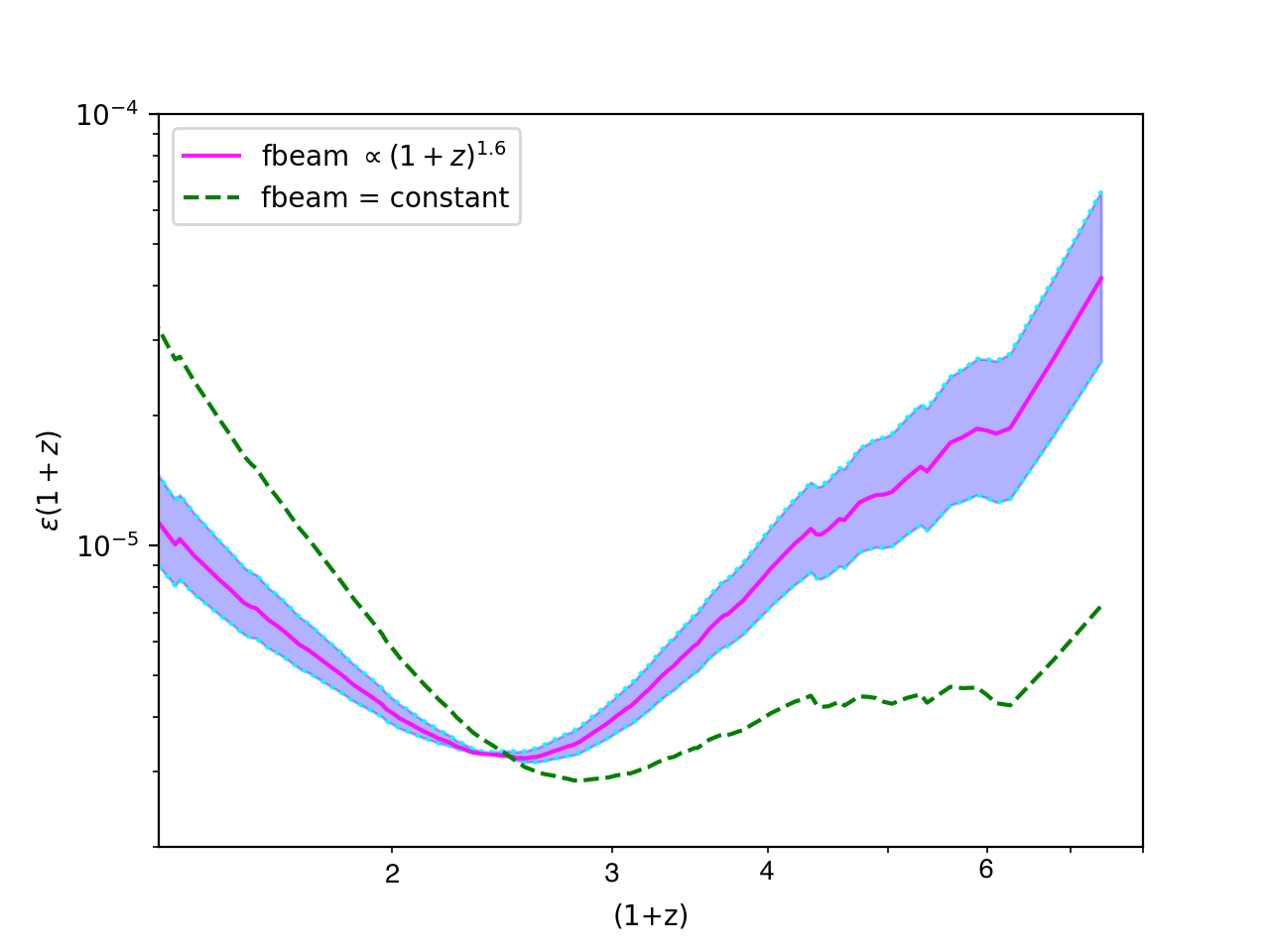}
    \caption{Fraction of stars $\epsilon(1+z)$ that make long gamma-ray bursts assuming the lGRB rate density directly traces the MD14 star formation rate density.  The magenta line and purple region show this quantity accounting for jet beaming angle evolution seen in the data $\theta_{j} \propto (1+z)^{-0.8 \pm 0.2}$.  The green dashed line shows $\epsilon(1+z)$ assuming no beaming angle evolution.}
    \label{fig:GRBfrac}
\end{figure}

\subsection{Estimating the Fraction of Stars Producing lGRBs}
  In our prescription above, we have assumed that some given fraction of stars (parameterized by the function $\epsilon(z)$) produces GRBs.  However, metallicity plays a strong role in stellar evolution, affecting the stellar structure, as well as the mass and angular momentum loss of a massive star  - quantities that are all crucially connected to whether or not a GRB will be successfully produced in its collapse.  And because metallicity evolves through cosmic time (e.g. \cite{Pet97,LL09,Yuan13}), we therefore might reasonably expect that the fraction of stars that produce GRBs will evolve with redshift, with more stars able to produce GRBs at lower metallicities (higher redshifts).  This is the motivation behind our parameterizations of $\epsilon(z)$ in Figure~\ref{fig:SFRevolv}, where we assumed a power-law evolution of the fraction of stars that produce GRBs.  We note that an important consideration in all of this is whether the star is in a binary system, and how this (along with metallicity) plays a role in the evolving fraction of stars that produce GRBs \citep{MT20}. 
  
   Therefore, another approach we may take in using the lGRB rate to learn something about star formation history, is to assume that the lGRB rate roughly traces the MD14 functional form of the global SFR, and solve for the fraction of stars that produce lGRBs.  In other words, one can take an assumed star formation rate, and - given the observed GRB rate - estimate the fraction of stars that make GRBs as a function of redshift:
  
  \begin{equation}
     \epsilon(z) = (\dot{dN}/dz)({\rm f_{beam}(z)})  \left(\frac{(1+z)}{dV/dz} \right) \frac{1}{\dot{\rho}_{\rm SFR}(z)}
  \end{equation}  
  
  \noindent Where we use
  \begin{equation}
  \dot{\rho}_{\rm SFR}(z)  = .0015\frac{(1+z)^{2.7}}{(1+[(1+z)/2.9]^{5.6})} \rm M_{\odot} yr^{-1} Mpc^{-3}. 
  \end{equation}
 \noindent for our star formation rate \citep{MD14}.
 
 We show this estimate for $\epsilon(z)$ in Figure~\ref{fig:GRBfrac}, where the magenta line (and purple region) indicates our estimate accounting for jet opening angle evolution and the green dashed line assumes no beaming angle evolution with redshift.  We choose to conservatively normalize the curves to a value of $\sim 5x10^{-6}$ at a redshift of $(1+z) \sim 3$, where star formation peaks.  We obtained this value by assuming roughly $0.1 \%$ of stars result in a supernova - of these supernovae, only about $\sim 15 \%$ \citep{Sm11} are of Type Ib/c, the type associated with long gamma-ray bursts.  Of this subset of Type Ib/c supernovae, only about $10\%$ \citep{Ch07, Kan13} successfully launch a GRB jet (due to conditions such as sufficient angular momentum and magnetic flux to launch a jet powerful enough to pierce through the progenitor envelope; a discussion of some of these issues can be found in \cite{LR19b}). This normalization is a big uncertainty, of course, and there is room for a range of values given our current state of knowledge. \\ 
 
 Regardless of the normalization, we can try to understand the resulting shape of the curves in Figure~\ref{fig:GRBfrac}.  There is a counter-intuitively large dip in the fraction of stars that make GRBs right at the peak of star formation, when we take this approach.  The increase in $\epsilon(z)$ at high redshifts may be plausible due to decreasing metallicity and possibly a top-heavy IMF at higher redshifts.  The increase in $\epsilon(z)$ at lower redshifts is uncertain and may, again, be a reflection of the breakdown between single star collapsar progenitors and lGRBs (see \S 3.3 above on the excess at low redshifts).  However, ultimately, the curve we find for $\epsilon(z)$ - under the assumption that the GRB rate traces the MD14 SFR - may be emphasizing that GRBs, in fact, do {\em not} trace the global star formation rate. 
 
  It is important to note, however, that the relative fraction of stars that produce lGRBs changes significantly {\em when accounting for beaming angle evolution of the GRB jet}.  As seen in Figure~\ref{fig:GRBfrac}, there is a much higher fraction of stars that make GRBs at high redshifts and relatively less at low redshift, when accounting for the change in average jet beaming angle over cosmic time.  Regardless of the validity of the underlying assumption of the lGRB rate tracing the global SFR, this emphasizes the importance of accounting for jet opening angle evolution when trying to understand the relationship of lGRBs to their progenitor systems.
  \\

\section{Conclusions}

 Observations suggest that the jet opening angles of lGRBs evolve over cosmic time, with lGRBs at higher redshifts more narrowly beamed than those at lower redshifts.  In this paper we have: 1) estimated the star formation rate from the gamma-ray burst formation rate, accounting for the evolution of the distribution of GRB jet opening angles (and given an assumption about the fraction of stars that make long gamma-ray bursts), 2) estimated the fraction of stars that make long gamma-ray bursts under the assumption that lGRBs trace the global star formation rate as parameterized by MD14.\\  

\noindent Our main results are as follows:
\begin{itemize}
    \item When accounting for beaming angle evolution - with lGRBs more narrowly beamed at higher redshifts - we find a higher relative star formation rate at high redshifts.  {\em Depending on the strength of the beaming angle evolution and the normalization of the inferred SFR, the SFR can be up to an order of magnitude higher than the canoncial MD14 estimate.}  Our inferred SFRs from the GRB rate may be indicating a specific metallicity dependent SFR (see, e.g., \cite{Bjor19,Chrus2020}), given the low-metallicity requirements for succesfully launching a GRB jet in a massive star. \\
   
    \item There appears to be an excess in our SFR estimates at {\em low redshifts} relative to the MD14 rate (again, depending on the normalization we choose).  Accounting for beaming angle evolution lessens this excess, which may suggest the importance of accounting for the evolution. Alternatively, this could be a reflection of the breakdown of a one-to-one correspondence between lGRBs and massive star progenitor systems at low redshifts.  In other words, if multiple systems (including binary merger systems) contribute significantly to the GRB rate at low redshifts this may lead to such an excess at low redshifts.   \\
    
    \item Under the assumption that GRBs trace the MD14 star formation rate, we estimate the fraction of stars that produce lGRBs (in order to be consistent with the observed GRB rate), once again accounting for beaming angle evolution.  Although the overall normalization of this curve is uncertain, we find that this approach implies a higher fraction of stars in the early universe produce GRBs.  This result is plausible in light of the fact that low metallicity conditions are conducive to launching a successful GRB. We also find, using this approach, that a higher fraction of stars produce GRBs at lower redshifts than at the peak of star formation (although less so when beaming angle evolution is accounted for).  As discussed above, this somewhat unexpected result could reflect the breakdown of a one-to-one correspondence between lGRBs and massive star progenitors at low redshifts, and may also indicate the implausibility of assuming that the lGRB rate density follows the SFR as parameterized by MD14.
\end{itemize}

  Because of the extreme luminosity of long gamma-ray bursts, they remain powerful probes of the early universe, and potentially important tools with which to measure the star formation rate at redshifts that are inaccessible by other methods.  That the jet opening angle of lGRBs may evolve over cosmic time, with jets in the early universe being more narrowly beamed than those at lower redshifts, has important implications on estimates of the star formation rate from the lGRB rate - implying it has perhaps up until now been largely underestimated.  As the next generation of telescopes is launched - including deep space optical and infrared probes such as the {\em James Webb Space Telescope} and  {\em Nancy Grace Roman Telescope,} as well as transient detectors such as {\em Theseus} and the {\em Space Variable Objects Monitor} - we will get a more extensive probe into the early universe.  In addition, new methods employing measurements of the neutrino flux \citep{Riya20}, for example, could enable us to more securely ascertain star formation during these epochs, allowing us to test our predictions of the SFR at high redshift, and gain a better understanding of the history of star formation throughout our Universe.\\

\section{Acknowledgements}
  We thank the referee for a very thoughtful report which led to many improvements in this manuscript.  We are very grateful to John Beacom for interesting discussions, and a number of helpful comments and suggestions related to this work.  We also thank Vahe' Petrosian for discussions on the rate at low redshifts.  This work was supported by the US Department of Energy through the Los Alamos National Laboratory.  Los Alamos National Laboratory is operated by Triad National Security, LLC, for the National Nuclear Security Administration of U.S. Department of Energy (Contract No. 89233218CNA000001). J. ~L. ~J. and A.~A. are  supported  by  a  LANL  LDRD Exploratory  Research  Grant  20170317ER.  LA-UR-20-23600
  
  \section{Data Availability}
  The data underlying this article are publicly available at \url{https://iopscience.iop.org/article/10.3847/1538-4357/ab0a86}. 

%%%%%%%%%%%%%%%%%%%%%%%%%%%%%%%%%%%%%%%%%%%%%%%%%%



\bibliographystyle{mnras}
\bibliography{refs} % if your bibtex file is called example.bib

%\begin{figure*}
%    \centering
%    \includegraphics[width=3.5in]{RedshifthistoALL.png}\includegraphics[width=3.5in]{RedshifthistoTHET.png} 
%    \caption{Redshift distributions of all GRBs (left) and those with jet opening angle measurements (left).  Roughly 1/3 of GRBs at all redshifts have opening angle measurements.}
%    \label{fig:my_label}
%\end{figure*}

% \begin{figure*}
%    \centering
%    \includegraphics{HighZSFRRenorm.png}
%    \caption{Star formation rate as a function of redshift assuming a constant fraction of stars make GRBs, but accounting for beaming angle evolution.  Curves are normalized to the Madau-Dickinson star formation rate at a redshift $(1+z) = 2$.}
 %   \label{fig:my_label}
%\end{figure*}
%%%%%%%%%%%%%

%%%%%%%%%%%%%%%%% APPENDICES %%%%%%%%%%%%%%%%%%%%%

%%%%%%%%%%%%%%%%%%%%%%%%%%%%%%%%%%%%%%%%%%%%%%%%%%

% Don't change these lines
\bsp	% typesetting comment
\label{lastpage}
\end{document}